\documentclass[useAMS,usenatbib,usegraphicx]{mn2e}

\newcommand{\htcn}{H$^{13}$CN}
\newcommand{\htnc}{HN$^{13}$C}
\newcommand{\hcn}{H$^{12}$CN}
\newcommand{\hnc}{HN$^{12}$C}

\newcommand{\cm}{cm$^{-1}$}
\newcommand{\teff}{T$_{\rm eff}$}

\newcommand{\HL}{H\"{o}nl--London}

\newcommand{\JMS}{J. Mol. Spect.\ }
\newcommand{\JCP}{J. Chem. Phys.\ }

\newcommand{\AAA}{A\&A}

\newcommand{\ApJ}{ApJ}

\newcommand{\leen}{lab/empirical energy level list}

\newcommand{\ai}{{\em ab initio}}

\newcommand{\dpr}{$^{\prime\prime}$}
\newcommand{\pr}{$^{\prime}$}

\title[A linelist for \htcn/\htnc]{A H$^{13}$CN/HN$^{13}$C linelist, model atmospheres and synthetic spectra for carbon stars.}

\author[Harris et al.]{G. J. Harris$^1$, F. C. Larner$^1$, J. Tennyson$^1$, 
\newauthor
B. M. Kaminsky$^3$, Ya. V. Pavlenko$^{3}$, and H. R. A. Jones$^4$\\
$^1$ Department of Physics and Astronomy, University College London, London, WC1E 6BT, UK.\\
$^2$ Physics Department, Imperial College London, South Kensington campus, London, SW7 2AZ. UK \\
$^3$ Main Astronomical Observatory, National Academy of Sciences, Zabolotnoho 27, Kyiv-127 03680, Ukraine.\\
$^4$ Centre for Astrophysics Research, University of Hertfordshire, Hatfield. AL10 9AB, UK. }

\begin{document} 

\maketitle
                                         
\begin{abstract}
A line list of vibration-rotation transitions for $^{13}$C substituted HCN is
presented. The line list is constructed using known experimental levels where
available, calculated levels and {\it ab initio} line intensities originally
calculated for the major isotopologue. Synthetic spectra are generated and
compared with observations for cool carbon star WZ Cas. It is suggested
that high resolution HCN spectra recorded near 14 $\mu$m should be 
particularly sensitive to the $^{13}$C to $^{12}$C ratio.

\end{abstract}


\section{Introduction.}
\label{sec:Intro}

Carbon giant stars are though to arise from the third dredge up of
asymptotic giant branch (AGB) stars, which pollutes the envelope and
atmosphere with nuclear processed material from the interior. This
increases the abundance of carbon and the $^{13}$C/$^{12}$C ratio to
well above the terrestrial level (0.011). In fact the
$^{13}$C/$^{12}$C ratio in carbon giants has been measured to be as
high as one-third \citep{ab97}. There are a number line lists
or opacity functions available for \hcn\
\citep{Jorgensen1,Jorgensen2,ao98,ha02b,ha06}. However, when
considering the contribution to opacity by molecular species such as
HCN it may not be sufficient to account only for \hcn, but also for
the isotopologue
\htcn.

The calculation of a complete triatomic linelist is computationally
expensive, requiring several tens of thousands of CPU hours
\citep{te07}. However, within the Born-Oppenheimer approximation the
electronic structure of isotopologues is identical. Thus both the
potential energy and electric dipole moment functions are identical
for all isotopologues. This implies that the vibration-rotation
frequencies and transition intensities are likely to be very similar.

For a hetronuclear molecule, such as HCN, the main differences between
the spectra of isotopologues are caused by the change in reduced
mass. In the harmonic approximation the vibrational contribution to
the line frequency is proportional to $\rho^{-\frac{1}{2}}$, where
$\rho=m_1m_2/(m_1+m_2)$ is the reduced mass. So the frequency of the
CN stretch vibrational mode of \hcn\ and \htcn\ will differ by the
order of 1\% and the bend and H-C stretch mode by less.  In terms of
both the observation of \htcn\ and the line blanketing of a model
atmosphere this shift in line frequency is more important than the
small differences in line intensity between comparable lines of \hcn\
and \htcn.

It is the aim of this work to compute a set of energy levels and line
frequencies for \htcn\ and \htnc.  These energy levels will be used in
conjunction with the Einstein A coefficients for \hcn\ and \hnc\
computed by \citet{ha02b}, to generate a \htcn/\htnc\ linelist. The
line frequencies in this 
\htcn/\htnc\ linelist have been corrected using available laboratory
data in order to increase its accuracy. The new linelist has been used
to compute synthetic spectra for C-stars, with different
$^{13}$C/$^{12}$C ratios using star WZ Cas as a prototype. 

\section{Construction of a \htcn\  and \htnc\  linelist.}
\label{sec:constlist}

\subsection{{\em ab initio} energy levels.}
\label{sec:enlev}


Using the existing HCN/HNC potential energy surface of \citet{vm01} a
set of {\em ab initio} rotation-vibration energy levels for \htcn\ and
\htnc\ were calculated for angular momenta of $J$ = 0, 1, 2, 3, 5, 10,
20, 30, 40, 60 and for both even ($e$) and odd ($f$) parity. The
states were computed up to an energy of at least $10000 +
B(J(J+1))$~\cm\ above the \htcn\ ground state, where $B\sim1.5$~\cm\ is
the HCN rotational constant.

The DVR3D suite of codes \citep{te04} was used for all these calculations. DVR3D uses an exact kinetic energy operator and discrete variable representation for the nuclear motion wavefunctions. We have used the initial basis set functions and parameters used by \citet{ha02b} for the computation of an extensive \hcn/\hnc\  linelist.
The basis functions are Legendre polynomials for the angular grid points and Morse oscillator-like functions for the radial grids.  
Jacobi coordinates were used, with 50 grid points for the angular coordinate, 35 grid points for the first ($R$) radial coordinate, 21 for the second ($r$) radial coordinate. Where $r$ is the distance from C to N nuclei and $R$ the distance from the H nucleus to the centre of mass of the C-N diatom.
The parameters for the Morse oscillator like basis functions in the r coordinate are $r_e = 2.3$ $a_0$, $D_e(r) = 29.0$ $E_h$ and $\omega_e(r) = 0.0105$ $E_h$ and $R_e = 3.2$ $a_0$, $D_e(R) = 5.0$ $E_h$ and $\omega_e(R) = 0.004$ $E_h$ for the R coordinate. Where $r_e$ is the equilibrium distance, $D_e$ is the dissociation energy and $\omega_e$ is the harmonic frequency.


\subsection{Vibrational assignments and rotational constants.}
\label{sec:rotcon}

Only the exact quantum numbers for a hetronuclear triatomic, angular
momentum ($J$) and parity are known for each state computed by
DVR3D. To determine the rotational constants for each vibrational
state it is first necessary to assign approximate vibrational quantum
numbers to each rotation-vibration state. The approximate vibrational
quantum numbers are C-H stretch ($v_1$), bend ($v_2$), C-N stretch
($v_3$), vibrational angular momentum ($l$) and isomer (\htcn\ or
\htnc ). Where $l$ take values in steps of 2 from 0 or 1, if $v_2$ is even or
odd respectively, up to the lower of $v_2$ or $J$. So
for a state with $v_2=5$ and $J\ge5$ then $l$ can take values of
1,3,5, but if $J=3$ or $J=4$ then $l$ can only take values of 1 or 3.

The quantum number assignments were made using a method similar to
that described in \citet{ha06}. Initially quantum numbers were
assigned to the \htcn\ $J=0$ purely vibrational states up to 10~000
\cm\ above the zero point energy. The first states to be assigned were
the lowest lying states of each of the 3 \htcn\ fundamental modes. The
vibrational expansion eq.~\ref{eq:vibexp} was then least squares fit
to the assigned states, which provided values for the fundamental
frequencies $\omega_i$.
\begin{eqnarray}
E(v,l)+E_0 & = & \sum^3_{i=1} \omega_i(v_i+d_i/2) \nonumber \\
           & + & \sum^3_{i=1}\sum^3_{j=1} x_{ij}(v_i+d_i/2)(v_j+d_j/2) \nonumber \\
           & + & x_ll^2
\label{eq:vibexp}
\end{eqnarray}

The fundamental frequencies allow estimates for the energy of more
highly excited states to be made, which aid further assignments. Next,
eq.~\ref{eq:vibexp} was least squares fit to the newly assigned states
so that the higher order vibrational constants $x_{ij}$ could also be
determined. This process was repeated until all $J=0$ \htcn\ states up
to 10~000~\cm\ were assigned. The $J$=0 \htnc\ states were fit
separately from the \htcn\ states, but the assignments made
simultaneously. The $J=1$ $e$ and$f$ states were assigned by
comparison with the $J=0$ states and by fitting equation
\ref{eq:vibexp} to verify the assignments.

In order to aid the assignments of states with $J>1$ we have least
squares fit the assigned levels for each vibrational state and each
parity with:
\begin{eqnarray}
E(v,J) & = & E(v)+B_v[J(J+1)+l^2] \nonumber \\
       &   & -D_v[J(J+1)+l^2]^2 + ... 
\label{eq:rotexp}
\end{eqnarray}
where $E(v)$ is the vibrational energy of the band, $B_v$ is the
rotational constant and $D_v$ is the centrifugal distortion
constant. This allows energies for the states with the next highest
$J$ to be estimated and assigned. For further verification these
assigned states were then fitted with the vibrational expansion.  This
process was carried out on an increasing $J$ by $J$ basis until all
the computed energy levels up to $J=60$ had been assigned. Once the
assignments were completed, a final fit of the rotational expansion
was performed for each vibrational state and parity. A total of 390
vibrational states of \htcn/\htnc\ have been studied all with a
vibrational energy ($E(v)$) less than 10~000~\cm . Using this final
set of rotational constants a set of calculated rotation-vibration
energy levels for \htcn\ and \htnc\ was computed for all values of $J$
up to 60.

At stellar temperatures transitions between levels with lower state
energy over 10~000~\cm\ do contribute significantly to opacity. These
lines tend to be weak and numerous so that they form an almost
continuous opacity source. In order to account for these high
temperature weak lines we have supplemented the \htcn/\htnc\ energy
level list with high energy H$^{12}$CN/HN$^{12}$C energy levels.


\subsection{Energy levels determined from laboratory data.}
\label{sec:expen}

Line frequencies determined in the laboratory are significantly more
accurate than than purely {\em ab initio} data.  Therefore, where
possible we have incorporated laboratory determined energy levels into
our linelist. \htcn\ has been well studied in the laboratory
\citep{le82,sm89,ma95,ma00,de03,de04}, but \htnc\ has been studied to
a far lesser extent \citep{ma01}. \htnc\ laboratory determined energy
levels have therefore not been incorporated in to the linelist.

\citet{ma00} have provided an electronic database of \htcn\  line frequencies, 
from this we have compiled a list of \htcn\ laboratory determined
energy levels, in the same way as \citet{ha06}. Many line frequencies
of the studied bands remain unmeasured, there are therefore missing
energy levels for many of the vibrational states. These missing energy
levels were interpolated by means of the rotational expansion to give
a complete list list of energy levels in $J$ for each band up to the
maximum $J$ measured by \citet{ma00}. As the rotational expansion is
divergent, it cannot be reliably used to extrapolate the laboratory
determined energy levels beyond the maximum $J$ measured by
\citet{ma00}. To extend the laboratory determined energy level list to
$J=60$, the maximum extent of the \citet{ha02b} linelist, we have used
a correction to the {\em ab inito} energy levels described in
\ref{sec:rotcon}. This constant is given by:
$C=E_{lab}(J_{max})-E_{ai}(J_{max})$, where $E_{lab}(J_{max})$ is the
energy of the state in the laboratory energy level list with the
highest angular momentum ($J_{max}$), and $E_{ai}(J_{max})$, is the
energy of the \ai\ energy level with $J=J_{max}$. This lab/empirical
energy level list contains 4425 energy levels, which have been
incorporated into the calculated list of \htcn/\htnc\ energy levels
described in section \ref{sec:rotcon}. The format of the energy level
data file is similar to that used by \citet{ha06}, a extract from the
file is given in table \ref{tab:energies}.


\begin{table*}
 \centering \begin{minipage}{140mm}
\caption{A sample from the \leen, which is available in full, in electronic form, from either the CDS archive (http://cdsweb.u-strasbg.fr/cgi-bin/qcat?/MNRAS/) or from our website (http://www.tampa.phys.ucl.ac.uk/ftp/astrodata).}
\begin{tabular}{rrrrrrrrrrrrrrr}
\hline

Index & $J$ & P & n & E$_{\rm ai}$ (\cm) & iso & $v_1$ & $v_2$ & $l$ & $v_3$ & E$_{\rm lab}$ (\cm) & error (\cm) & label1 & label2 \\ \hline
  1778  & 3 & 1   & 28  & 4867.640942  & 0  & 0  & 4  & 2  & 1  & 4862.194989   & 3.08E-02 &  c &  I \\
  1779  & 3 & 1   & 29  & 5219.561016  & 1  & 0  & 0  & 0  & 0  &               &          &    &  I \\
  1780  & 3 & 1   & 30  & 5351.139192  & 0  & 1  & 3  & 1  & 0  & 5350.627028   & 6.08E-04 &  e &  I \\
  1781  & 3 & 1   & 31  & 5361.140796  & 0  & 1  & 0  & 0  & 1  & 5360.707912   & 3.00E-04 &  e &  I \\
  1782  & 3 & 1   & 32  & 5384.899692  & 0  & 1  & 3  & 3  & 0  & 5380.086981   & 7.55E-04 &  e &  I \\
  1783  & 3 & 1   & 33  & 5516.043520  & 0  & 0  & 8  & 0  & 0  &               &          &    &  I \\ \hline

\end{tabular}
\label{tab:energies}
\end{minipage}
\end{table*}

In table \ref{tab:energies} the column labelled index is the index
number given to the corresponding \hcn\ energy level by \citet{ha02b},
$J$ and P are the exact quantum numbers of angular momentum and
parity, n is the number of the energy level in the $J$-P symmetry
block, E$_{ai}$ is the value of the calculated energy level, iso
labels the state as either \htcn\ (iso=0) or \htnc\ (iso=1), $v_1$,
$v_2$, $l$, and $v_3$ are the approximate quantum numbers, E$_{\rm
lab}$ is the lab/empirical energy, label1 is a single character label
which is either 'e' for a laboratory determined energy level, 'c' for
an interpolated energy level or 't' for a corrected \ai\ energy level,
finally label2 is a single character label which identifies the
calculated energy as either I for an interpolation made using the
fitted \htcn/\htnc\ rotational constants or C an \ai\ \hcn/\hnc\
energy from the \citet{ha02b} energy level list. The final energy
level list is organised into symmetry blocks according to the $J, P$
quantum numbers.

The error on the energy level takes three forms, for a lab determined
energy level this is the compound error of the line frequency
measurements used to derive the energy level. For an energy level
computed by interpolation of laboratory data this is the standard
deviation on the fit of eq. \ref{eq:rotexp}. For a corrected \ai\
energy level the error is the difference between the energy predicted
by the fit of eq. \ref{eq:rotexp}, to lab determined energy levels,
and the corrected calculated \htcn\ or \ai\ \hcn\ energy.

\subsection{Partition function.}
\label{sec:part}

It is essential to know the temperature dependent rotation vibration
partition function to compute line intensities at thermodynamic
equilibrium. The rotation-vibration partition function of \htcn/\htnc\
was calculated by direct summation over all the energy levels in the
energy level list, using:
\begin{equation}
Q_{rv}(T) = \sum_i (2J_i+1) \exp\left(\frac{-E_i}{kT}\right)
\end{equation}
where $J$ is the angular momentum quantum number of the state $i$,
$E_i$ the energy of the state, $k$ is the Boltzmann constant and $T$
is temperature. For temperatures between 500 and 10~000~K the
partition function was least squares fit with the function:
\begin{equation}
\log(Q_rv(T)) = \sum_{i=0}^n a_i (log_{10}(T))^i
\end{equation}
with $n=4$. The standard deviation from $\log_{10}(Q(T))$ is 0.0075
and the coefficients are $a_0=-57.453223$, $a_1=86.387042$,
$a_2=-46.668113$, $a_3=11.018178$ and $a_4=-0.93943933$. The summation
used laboratory determined energy levels where available and in
preference to the calculated \htcn\ energy levels. The calculation was
augmented with \hcn\ energy levels for vibrational energies greater
than 10~000~\cm. At 296~K the partition function was calculated to be
152.93. This can be compared with a value of 148.72 calculated at 296~K
for \hcn\ using a similar procedure \citep{ba02}.

\subsection{Einstein A coefficients and Laboratory determined band dipoles.}
\label{sec:banddip}

The intensities of a few bands of \htcn\ and \htnc\ have been measured
\citep{sm89,ma95,de03,de04}, these are the stretching fundamentals,
the first CN stretch $v_3$ hot band, the $2v_1$, $2v_1+v_3$ and $2v_2$
overtone bands. The laboratory determined band dipoles of \htcn\ and
\hcn, together with {\em ab initio} band dipoles for \hcn\ calculated
by \citet{ha02a,ha02b} are listed in table \ref{tab:dipole}. The band
dipoles for \htcn\ match those of \hcn\ to within 15\%, except for the
$v_3$ fundimental and its hot band. This verifies that the \hcn\
Einstien A coefficents are a reasonable approxomation to those of
\htcn.  The $v_3$ bands for both \hcn\ and \htcn\ is unusually weak
and has an intensity structure which shows an unusual double peak in
the R branch. These occur at slightly different $J$\dpr\ for each
isotopologue. To more accurately account for the unusual intensity
structure of the $v_3$ fundamental in the \htcn\ linelist, we have
used the band dipoles and Herman-Wallace constants given by
\citet{ma95} to compute Einstein A coefficients for lines of the CN
stretch fundamental and hot bands. Throughout this work these lab
determined Einstein A coefficients are substituted for the \ai\
Einstein A coefficients of \citet{ha02b}. Einstein A coefficients for
individual rotation vibration lines are calculated from the band
dipole and Herman-Wallace constants with the following formula.
\begin{equation}
\label{eq:mutoA}
A_if = C\nu^3\frac{F_{HL}F_{HW}}{(2J^{\prime}+1)}\mu^2
\end{equation}
where $F_{HL}$ and $F_{HW}$ are the \HL\ and Herman-Wallace factors as
described by \citet{ma95}, $\mu$ is the band dipole, $\nu$ is line
frequency, and $C=64\pi^4/(3c^3h)$. To return an Einstein A
coefficient in $s^{-1}$ with dipole in Debye and frequency in
cm$^{-1}$, then $C$ should be set to $3.136186 \times 10^{-7}$.

\begin{table*}
 \centering
 \begin{minipage}{140mm}
\caption{Available laboratory determined band dipoles (Debye) for \htcn, the corresponding theoretical and laboratory data for \hcn\  are shown for comparison. Fitting errors in the last quoted digit are given in parentheses, where available.}
\begin{tabular}{ccccc}
\hline
$(v_1^{\prime},v_2^{\prime},l^{\prime},v_3^{\prime})$ &
$(v_1^{\prime\prime},v_2^{\prime\prime},l^{\prime\prime},v_3^{\prime\prime})$
& \htcn & \hcn & \hcn\ theory.  \\ \hline

 (0,2,0,0)  & (0,0,0,0) & 0.047(4)    & 0.0496(2)   & 0.0479(11) \\
 (0,0,0,1)  & (0,0,0,0) & 0.000309(2) & 0.001362(4)    & \\
 (0,1,1,1)  & (0,1,1,0) & 0.00263(3)  & 0.001794(3)    & \\
 (1,0,0,0)  & (0,0,0,0) & 0.085(5)    & 0.0831(17)  & 0.0853(16)  \\
 (2,0,0,0)  & (0,0,0,0) & 0.00731     & 0.00881(12) & 0.0086(4)   \\
 (2,0,0,1)  & (0,0,0,0) & 0.00065     & 0.00068(1)  & 0.000677(4) \\ \hline
\end{tabular}

\label{tab:dipole}
\end{minipage}
\end{table*}

\section{The new linelist}
\label{sec:linelist}

Using the \citet{ha02b} Einstein A coefficients, the \citet{ma95}
laboratory intensity measurements, the lab/empirical, computed and
\ai\ \hcn/\hnc\ energy levels, we have generated a new \htcn/\htnc\
linelist. In this linelist the weak transitions between high energy
states with vibrational energy greater than 10~000~\cm\ are are
accounted for by using only the \citet{ha02b} \ai\ data. For the
majority of transitions between states of lower energy the calculated
\htcn/\htnc\ energy levels  are used to
give line frequency and the \ai\ Einstein A coefficients of
\citet{ha02b} to calculate the intensity. However, where available, the
lab/empirical \htcn/\htnc\ energy levels (see section \ref{sec:expen}
) are used to calculate line frequency in preference to the calculated
energy levels. For the special cases of the $v_3$ fundamental and
first hot band we have used the band dipoles and Herman-Wallace
constants of \citet{ma95} to compute Einstein A coefficients and
intensity for individual lines.

We have truncated the \htcn/\htnc\ linelist at a minimum intensity of
$3\times10^{-28}$ cm~molecule$^{-1}$ at 3000 K. This results in a
linelist of 34.1 million lines which accounts for more than 99.9\%\ of
the opacity of the full linelist at 3000 K. The format of the linelist
is identical to that of \citet{ha06} and a sample of the linelist is
given in table \ref{tab:astrolist}, here $\nu$ is frequency, E\dpr\ is
lower state energy, A$_{if}$ is the Einstein A coefficient, $J$\dpr\
and $J$\pr\ are lower state and upper state angular momentum quantum
numbers, p is parity where 1 is even and 0 odd, n is the number of the
level in the $J$-parity block, index is the unique label of the energy
level, iso labels a \htcn\ state if 0 or a \htnc\ state if 1, $v_1$,
$v_2$, $l$, $v_3$ are the approximate quantum numbers. Where the
approximate quantum numbers have not been assigned a value of $-2$ is
given.  The full version of the linelist is available from either the
CDS archive http://cdsweb.u-strasbg.fr/cgi-bin/qcat?/MNRAS/ or from
our website http://www.tampa.phys.ucl.ac.uk/ftp/astrodata. Note these
site provide the data in the form of table \ref{tab:astrolist} or alternatively
as a file whose three columns are the wavelength in \AA, $g_{n}f_{nm}, n >
m$ for absorption and
the lower state energy, $E''$, eV.

The untruncated linelist can be obtained by downloading the original
set of \citet{ha02b} Einstein A coefficients, the new assigned energy
level list and running the supplied FORTRAN utility program
dpsort-H13CN-v2.1.f90. The Einstein A coefficient file from
\citet{ha02b} is sorted using \ai\ \hcn/\hnc\ frequencies, so that the
generated \htcn/\htnc\ linelist will no longer be in full frequency
order.

\begin{table*}
 \centering
 \begin{minipage}{180mm}
\caption{A sample from the \htcn/\htnc\  linelist, the strongest 34.1 million of which are available in electronic form, from either the CDS archive http://cdsweb.u-strasbg.fr/cgi-bin/qcat?/MNRAS/ or from our website http://www.tampa.phys.ucl.ac.uk. Quantum numbers are: $J$ rotational level, $p$ parity, $n$ state number in $J,p$ block, $(v_1,v_2^\ell,v3)$ vibrational labels. Upper state denoted by ' and lower state by ''. ind gives the unique index number of each state; iso gives the isomer (0 =  \htcn, 1 = \htnc); lbl is ai for {\it ab initio} calculated frequency and lb for laboratory determined frequency.}

\begin{tabular}{cccccccccccccccccccccc}
\hline
$\nu$(cm$^{-1}$) & $E''$(cm$^{-1}$) & $J''$& $p''$ & $n''$ & $J'$& $p'$ & $n'$ & $A$(s$^{-1}$) & $i''$ & $i'$ 
& iso'' & $v_1''$ &$v_2''$ & $\ell''$ & $v_3''$& iso' &$v_1'$ &$v_2'$ & $\ell'$ & $v_3'$ &lbl\\
\hline
0.508814 & 10443.914579 & 39 & 1 & 141 & 39 & 0 & 109 & 2.379E-04 & 127951 & 129329 & 0 & 0 & 9 & 1 & 1 & 0 & 0 & 12 & 2 & 0 & ai \\
0.511822 &  7484.879768 & 42 & 0 &  23 & 42 & 1 &  33 & 8.824E-09 & 136763 & 135483 & 0 & 0 & 4 & 4 & 1 & 0 & 0 &  4 & 4 & 1 & lb \\
0.512459 &  8544.892933 & 12 & 0 & 116 & 12 & 1 & 153 & 2.261E-08 &  31656 &  29543 & 0 & 0 & 9 & 5 & 1 & 0 & 0 &  6 & 6 & 2 & ai \\
0.512850 &  4906.391346 &  8 & 1 &  29 &  8 & 0 &  20 & 5.828E-09 &  14639 &  16500 & 0 & 0 & 1 & 1 & 2 & 0 & 0 &  1 & 1 & 2 & lb \\
0.513045 &  5741.587834 & 32 & 0 &  16 & 32 & 1 &  24 & 7.630E-09 & 108856 & 107154 & 0 & 0 & 6 & 4 & 0 & 0 & 0 &  6 & 4 & 0 & lb \\ \hline
\end{tabular}

\label{tab:astrolist}
\end{minipage}
\end{table*}

\begin{figure*}
\includegraphics[angle=-90,width=84mm]{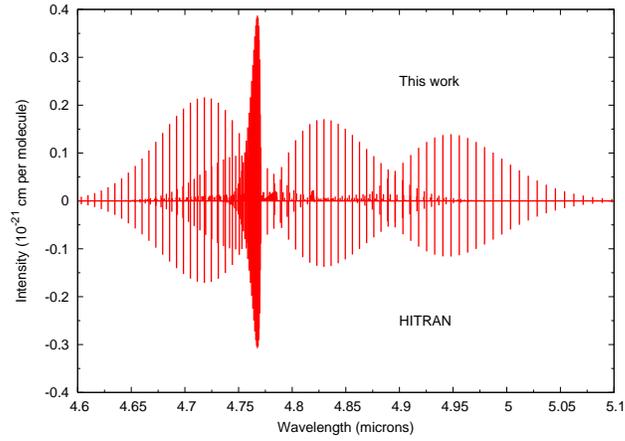}
\caption{The integrated intensity and frequency of lines of the \htcn\  $v_3$ (CN stretch) fundamental and the $3v_2$ (bend) overtone at 296~K. The lines from the linelist of this work are shown with positive intensity and lines from the HITRAN database \citep{ro05} are shown with negatively intensity.}
\label{fig:stick}
\end{figure*}

Figure \ref{fig:stick} shows the intensity and frequency of the \htcn\
lines from the HITRAN04 database \citep{ro05} against those from our
new \htcn\ linelist, for the $v_3$ (CN stretch) fundamentals and
$3v_2$ (bend) overtone. There is very good agreement between line
frequencies and good agreement between line intensities. For the
$3v_2$ overtone our intensities are around 20\% stronger than those
from HITRAN, the agreement between intensities is better for the $v_3$
fundamental.

\begin{figure*}
\includegraphics[angle=-90,width=84mm]{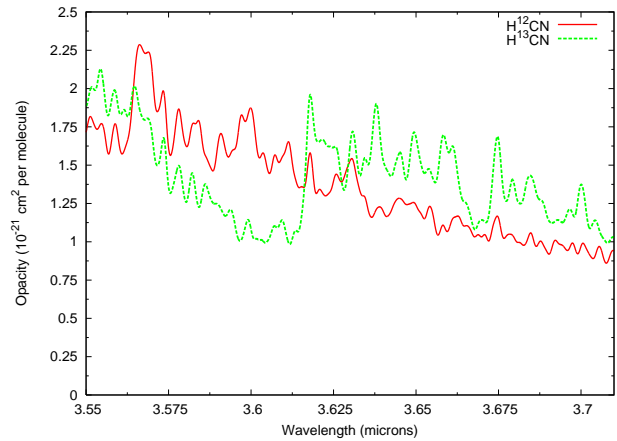}
\caption{The absorption cross-section per molecule for 
the \hcn\  and \htcn\  $v_2+v_3$ (bend and CN stretch) combination bands at 
3000~K.}
\label{fig:opacv2v3}
\end{figure*}

\begin{figure*}
\includegraphics[angle=-90,width=84mm]{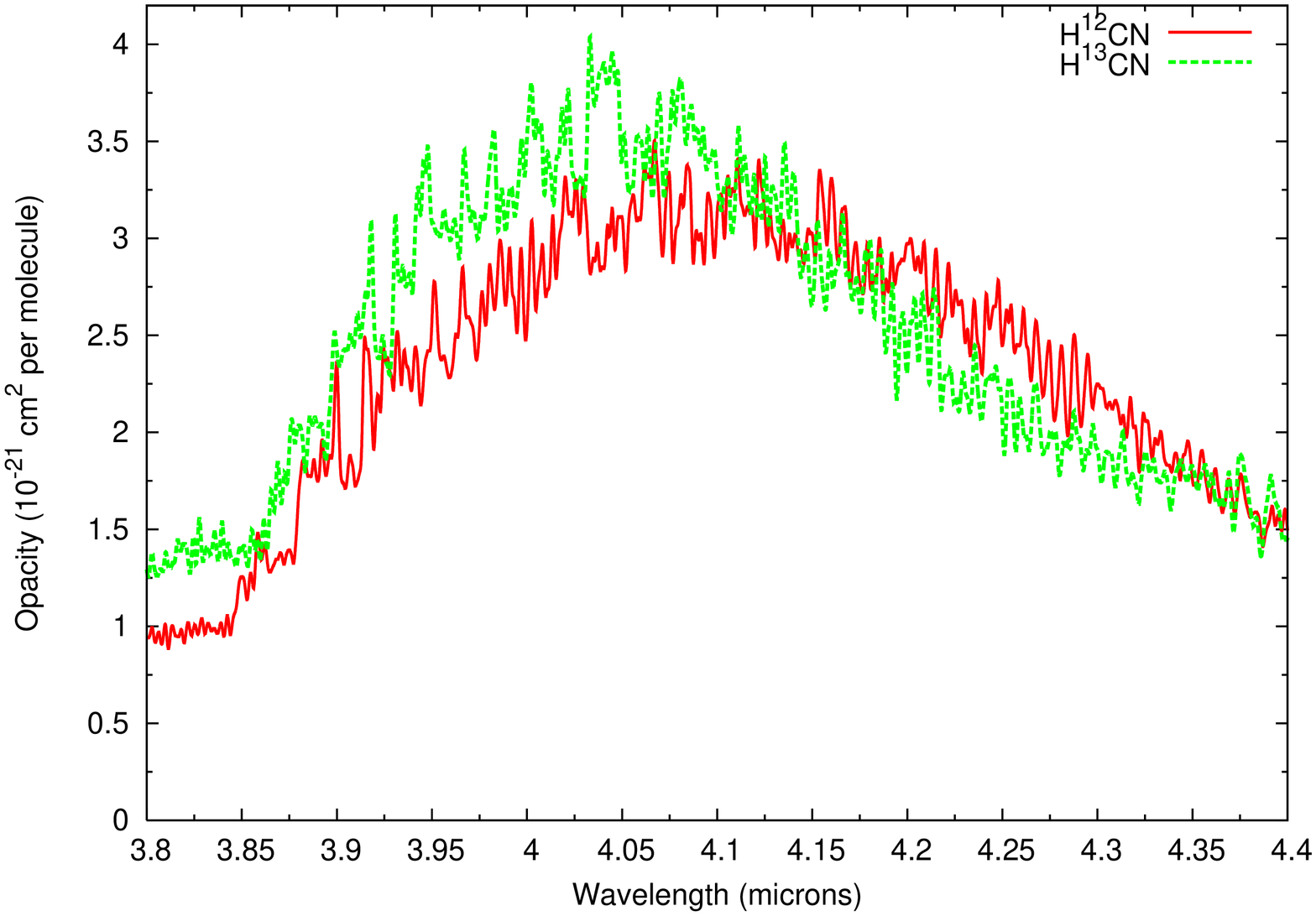}
\caption{The absorption cross-section per molecule for the \hcn\  and \htcn\  $v_1-v_2$ (bend and HC stretch) combination bands at 3000~K.}
\label{fig:opacv1v2}
\end{figure*}

\begin{figure*}
\includegraphics[angle=-90,width=84mm]{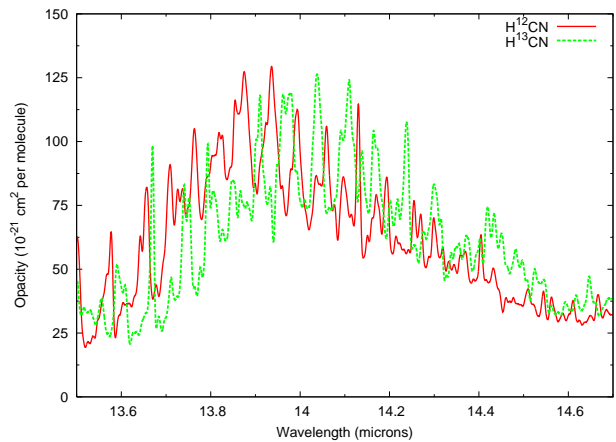}
\caption{The absorption cross-section per molecule for the \hcn\  and \htcn\  
$v_2$ (bend) fundamental and hot bands at 3000~K.}
\label{fig:opacv2}
\end{figure*}

Figures \ref{fig:opacv2v3}, \ref{fig:opacv1v2} and \ref{fig:opacv2}
show the opacity of \htcn, determined in this work, compared to that
of \hcn\ computed by \citet{ha06} at a temperature of 3000~K. The
opacity has been calculated by convolving the individual lines with
Gaussians of half-width at half-maximum of $\nu/4000$, where $\nu$ is
the wavenumber of the line, this gives a resolution comparable to that
achieved with the SWS spectrometer on the ISO satellite. The bands
shown in these figures all show strong Q-branches, which makes the
identification of some bands possible at high temperature and low
resolution. The identification of absorption by \htcn\ from the
$v_1-v_2$ bands is likely to be made difficult by the strong
background of \hcn. However, the $v_2$ and $v_2+v_3$ bands look to be
more promising candidate for identification in a C-star. This is
because the opacity of \htcn\ at the peak of the Q-branches of these
bands is up to 70\% stronger than the \hcn\ opacity, making them more
easily identifiable against a strong background of \hcn\ absorption.

\section{Synthetic and observed spectra for \protect{WZ~Cas}.}

\begin{figure*}
\includegraphics[angle=-90,width=84mm]{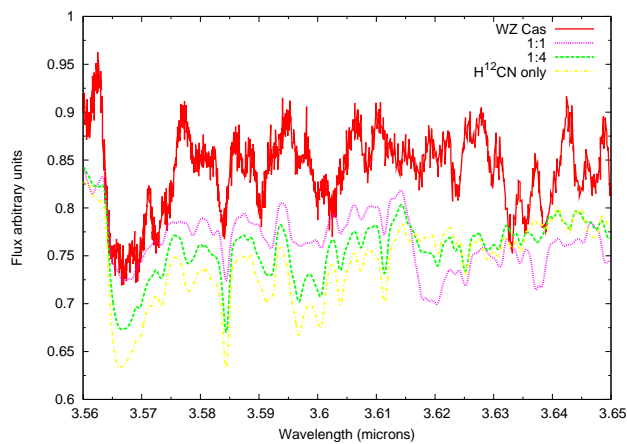}
\caption{Synthetic and observed spectra for WZ~Cas, the synthetic spectra have been displaced form the observed spectrum by 0.06 flux units. The synthetic spectra are shown for \htcn\  to \hcn\  ratios of 0, 1/4 and 1.}
\label{fig:wzcas}
\end{figure*}

Synthetic spectra for the carbon star WZ~Cas were calculated with the WITA6 
program \citep{Pavlenko2000}.
In addition to the same approximations, opacities and input parameters 
as in our earlier calculations \citep{ha06}, we have 
also included our new \htcn/\htnc\  opacity. These synthetic spectra 
are calculated using the best fit model atmosphere for WZ~Cas 
from \citet{ha06}, which has \teff $=$ 2800~K, 
$\log({\rm N_C}/{\rm N_O}) = -0.003$, $\log(g) = 0.0$. Synthetic spectra 
computed with \htcn/\hcn\  ratios of 0, 1/4 and 1 compared to the ISO/SWS 
observed spectrum of WZ~Cas \citep{ao98} are shown in figure 
\ref{fig:wzcas}. The effect of \htcn\  absorption can be seen in the synthetic 
spectra with \htcn/\hcn\ ratios of 1/4 and 1, however below a ratio of
1/4 it becomes difficult to identify \htcn\ absorption on the strong
background of \hcn\ lines. Although, the strength of the some of the
\htcn\ features in the theoretical spectrum indicates a useful
$^{12}$C/$^{13}$C ratio maybe measured around 3.62 and 3.64 $\mu$m there
remain other unidentified opacities in this region, e.g., the strong
observed bandhead at 3.63 $\mu$m.

\begin{figure*}
\includegraphics[width=84mm,angle=0]{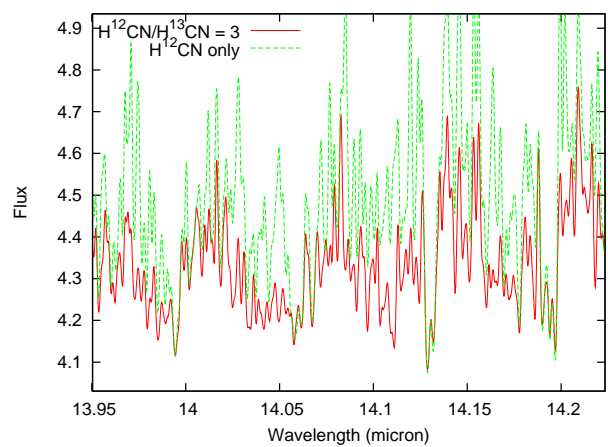}
\caption{Synthetic spectra computed of WZ Cas computed for the pure \hcn\
line list (dashed line) and \hcn\ +
\htcn\ lists. In the last case $^{12}C$/$^{13}C$ = 3 was adopted. 
The spectra are convolved to correspond to a resolution of 10000.} 

\label{fig:_30}
\end{figure*}

Another interesting spectral region to study HCN features we find
around 14 $\mu$m (see \cite{ha06}) for more details.  The predominant
features in this region are the Q branches of the and ($\Delta v_2=1$)
bands. Computed spectra for the 2800/0.0 model atmosphere given
by \citet{ha06} are shown in
Fig. \ref{fig:_30}. To simplify presentation of our results only a
short spectral range is shown here.  As we see from the comparison of
computed spectra with and without a contribution from \htnc\ they
differ significantly, at least at 14.05, 14.11, 14.23 $\mu$m, where
the strong features created by \htcn\ bands are located. This
region can be considered as very promising for the future
investigations of the carbon isotopic ratios in atmospheres of carbon
stars.
 
We have searched the ISO archive for suitable high resolution data.
The available data in the 14 $\mu$m regime for potentially 
suitable targets IRAS 15194-5115, RY Dra, T LYR, WZ CasA, Y Cvn covers a 
relatively large parameter space of stellar properties with high 
$^{12}$C/$^{13}$C ratio although none of the 
available spectra offer a combination of 
resolution, signal-to-noise and other features which we understood 
well enough to allow us to convincingly identify \htcn\ or \htnc\ 
features. Such analysis will have to await both better spectra and a
better characterisation of the other absorbing 
species in the same wavelength region.

\section{Conclusion.}
\label{sec:conc}

We present a new set of {\em ab initio} 
rotation-vibration energy levels for \htcn\ and \htnc\  which were 
calculated for angular momenta of 
$J$ = 0, 1, 2, 3, 5, 10, 20, 30, 40, 60 and for both even ($e$) and odd 
($f$) parity. 
The states were computed up to an energy of at least 
$10000 + B(J(J+1))$~\cm\  above the \htcn\ ground state, where
$B\sim1.5$~\cm\ is the HCN rotational constant.
The quantum number assignments were made using a method 
similar to that described in \citet{ha06}.

The new linelist has been incorporated into our computations of C-rich 
synthetic spectra. The detailed analysis of the infrared spectra
of C-giant star with high
$^{13}$C/$^{12}$C ratios ought to take account of \htcn\ and \htnc\ species.
Moreover in many cases HCN spectra probably provides the best chance of
determining the $^{13}$C/$^{12}$C ratios in atmospheres of
the coolest stars as the CO bands at 2.3 $\mu$m are
usually saturated
and other molecular bands are severely blended.
Our upgraded opacity sources can be used for 
the determination of carbon isotopic ratios in atmospheres of carbon stars. 
The most promising  regions are those around 3.6 and 14 $\mu$m where the 
the  $v_2+v_3$ (bend and CN stretch) combination bands
and $v_2$ (bend) fundamental and hot bands of \hcn\  and \htcn\ molecules
are located, respectively. However, to do this requires to use the 
spectral data
of the proper quality. 

Finally, the use of \hcn\ and \htcn\ lines for numerical
analysis of infrared spectra of evolved stars is restricted by the
incompleteness of presently available opacity sets, see
\citet{te07}. Special attention needs be paid to the computation of
other lines for polyatomic molecules such as C$_3$, NH$_3$, CH$_4$ and
C$_2$H$_2$, and their isotopologues.

\section*{Acknowledgements}

GJH thanks the UK Particle Physics and Astronomy Research Council (PPARC), 
for post-doctoral funding. The work YP is partially supported by the Recearch
Grant of Royal Society and the program ``Microcosmophysics'' 
of the National Space Agency and Academy of Sciences of Ukraine. 
The manipulation and analysis of the HCN/HNC linelist was carried out 
on the Keter computer facility of the HiPerSPACE computing centre at 
UCL which is part funded by PPARC. We thank PPARC/STFC for visitor grants.

\end{document}